\documentclass[a4paper]{jpconf}
\usepackage{graphicx}
\begin{document}
\title{Non-Newtonian gravity in finite nuclei}

\author{Jun Xu}
\address{Department of Physics and Astronomy, Texas A$\&$M
University-Commerce, Commerce, Texas 75429-3011, USA}
\ead{Jun.Xu@tamuc.edu}

\author{Bao-An Li}
\address{Department of Physics and Astronomy, Texas A$\&$M
University-Commerce, Commerce, Texas 75429-3011, USA}
\address{School of Science, Xi'an Jiao Tong
University, Xi'an 710049, P.R. China}
\ead{Bao-An.Li@tamuc.edu}

\author{Lie-Wen Chen}
\address{INPAC, Department of Physics and Shanghai Key
Laboratory for Particle Physics and Cosmology, Shanghai Jiao Tong
University, Shanghai 200240, China}
\address{Center of
Theoretical Nuclear Physics, National Laboratory of Heavy Ion
Accelerator, Lanzhou 730000, China}
\ead{lwchen@sjtu.edu.cn}

\author{Hao Zheng}
\address{INPAC, Department of Physics and Shanghai Key
Laboratory for Particle Physics and Cosmology, Shanghai Jiao Tong
University, Shanghai 200240, China} \ead{zh-i@sjtu.edu.cn}

\begin{abstract}
In this talk, we report our recent study of constraining the
non-Newtonian gravity at femtometer scale. We incorporate the
Yukawa-type non-Newtonian gravitational potential consistently to
the Skyrme functional form using the exact treatment for the direct
contribution and density-matrix expansion method for the exchange
contribution. The effects from the non-Newtonian potential on finite
nuclei properties are then studied together with a well-tested
Skyrme force. Assuming that the framework without non-Newtonian
gravity can explain the binding energies and charge radii of medium
to heavy nuclei within $2\%$ error, we set an upper limit for the
strength of the non-Newtonian gravitational potential at femtometer
scale.
\end{abstract}

\section{Introduction}

How to understand the nature of the gravitational force and unify it
with other fundamental forces are among the most important questions
to answer in the new century~\cite{11questions}. It has gradually
been realized that the traditional Newtonian gravitational
potential, which obeys the Inverse-Square-Law (ISL), can not explain
all the phenomena relative to gravity in nature. Great efforts have
been devoted to search for the possible existence of non-Newtonian
gravity that violates the
ISL~\cite{Fis99,Hoyle01,Adel03,Uzan03,Rey05,Kam08,New09}.
Especially, these studies have put upper limits of the strength of
the non-Newtonian gravity from as large as galaxy scale to as small
as 10 fm~\cite{Kam08}. At galaxy scale, various modified gravity
theories are used to study the non-Newtonian gravity such as the
scalar-tensor-vector gravity and $f(R)$ gravity. At short distances,
the origin of the non-Newtonian potential can be attributed to the
possible existence of extra dimensions within string/M theories or
exchanging light and weakly-coupled new particles in the
supersymmetric extension of the Standard Model.

The following form of the gravitational potential between two
objects of masses $m_1$ and $m_2$ at positions $\vec{r}_1$ and
$\vec{r}_2$ has been widely used at both galaxy scale and short
distances~\cite{Fuj71}
\begin{equation}\label{Fu}
V_{\rm grav}(\vec{r}_1,\vec{r}_2) =
-\frac{Gm_1m_2}{|\vec{r}_1-\vec{r}_2|}(1+\alpha
e^{-|\vec{r}_1-\vec{r}_2|/\lambda}).
\end{equation}
In the above, the first term is the traditional Newtonian potential
where $G$ is gravitational constant, while the second term
represents the non-Newtonian potential, where $\alpha$ and $\lambda$
are the strength and length scale parameters, respectively.

At galaxy scale ($\lambda \sim 10^{21}$ m), this simple Yukawa-type
non-Newtonian gravitational potential can be viewed as a reduced
form of the scalar-tensor-vector gravity or the $f(R)$ gravity in
the weak-field limit~\cite{Wof06,Cap09}, and it has been used to
successfully explain both the flattening of the galaxy rotation
curves away from the Kepler limit~\cite{San84,Mof96} and the Bullet
Cluster 1E0657-558 observations in the absence of dark
matter~\cite{Bro07}. For a recent review of applying this
Yukawa-type non-Newtonian gravitational potential on well-known
observations at galaxy scale, we refer the readers to
Ref.~\cite{Nap12}.

At short distances, the Yukawa potential in Eq.~(\ref{Fu}) may come
from exchanging a light and weakly-coupled spin-0 axion~\cite{Wil78}
or spin-1 U-boson~\cite{Fay80} corresponding respectively to an
attractive or repulsive potential. The exchanged boson is related to
several interesting new phenomena in particle physics and
cosmology~\cite{Fay07,Fay09} and may mediate the annihilation of
dark matter particles~\cite{Boe04a,Boe04b} which accounts for the
511 keV $\gamma$-ray from the galactic bulge as well.

Especially, the properties of a neutron star would be affected by
the possible existence of the non-Newtonian gravitational potential
if its length scale is smaller than the size of the star. In this
case, if one still wants to rely on solving the
Tolman-Oppenheimer-Volkoff using the equation of state (EOS) of
neutron star matter, the above Yukawa-type non-Newtonian
gravitational potential can be viewed as an extra potential between
nucleons in addition to the basic nucleon-nucleon interaction. It
has been found that the possible existence of the repulsive Yukawa
potential from exchanging a spin-1 U-boson would stiffen the EOS of
neutron star matter compared to that from a nucleon-nucleon
interaction only and thus increase both the mass and the radius of a
neutron star~\cite{Kri09}. Even if with a supersoft symmetry energy
at suprasaturation densities, the repulsive Yukawa potential could
still maintain a stable neutron star~\cite{Wen09}. In addition, the
core-crust transition density/pressure could also be affected and
the neutron star structure could be modified in the presence of the
Yukawa potential~\cite{Zhe11}. The coupling of U-boson and nucleon
has been successfully introduced into the relativistic mean-field
model and similar effects have been observed~\cite{Zhang11}.
Furthermore, it is shown that although the presence of hyperons and
quarks would soften the EOS of neutron star matter, the repulsive
Yukawa potential can stiffen the EOS and account for the heavy mass
of PSR J1614-2230~\cite{Wen11}. Thus, our knowledge of the
non-Newtonian potential and the nuclear EOS are both important in
understanding many interesting phenomena in nuclear astrophysics.

At a certain length scale $\lambda$, the non-Newtonian potential is
generally compared with the typical interaction for the
corresponding length scale, through which the strength of the
non-Newtonian potential is constrained if the interaction is well
determined. As mentioned above, the study and the constraint of the
non-Newtonian potential below 10 fm are still lacking. The most
typical system at femtometer scale is a finite nucleus, where the
typical interaction is the strong interaction. Thanks to the great
efforts made by the nuclear physicists, the properties of finite
nuclei can be very well explained by effective nucleon-nucleon
interactions. Especially, from the empirical values of isoscalar and
isovector macroscopic quantities, the parameters in the Skyrme
effective interaction can be inversely determined, leading to the
MSL0 Skyrme force~\cite{Che10}. This gives us the opportunity to
study the possible existence of the Yukawa-type non-Newtonian
potential in the finite nuclei whose properties have been well
described by the MSL0 Skyrme force, and constrain the strength of
the non-Newtonian potential at femtometer scale.

\section{Incorporating the Yukawa potential to the Skyrme-Hartree-Fock calculation}
\label{theory}

In the one-boson-exchange picture, the Yukawa-type non-Newtonian
gravitational potential between two nucleons at position $\vec{r}_1$
and $\vec{r}_2$ is written as
\begin{equation}\label{Vy}
V_{\rm Y}(\vec{r}_1,\vec{r}_2) = \pm \frac{g^2}{4\pi} \frac{e^{-\mu
|\vec{r}_1-\vec{r}_2|}}{|\vec{r}_1-\vec{r}_2|}.
\end{equation}
In the above, $g$ is the boson-nucleon coupling constant and it can
been expressed as $g=\sqrt{4\pi|\alpha| G m^2}$ comparing
Eq.~(\ref{Vy}) with Eq.~(\ref{Fu}), where $m$ is the nucleon mass.
$\mu=1/\lambda$ is the mass of axion or U-boson which is largely
quite uncertain, and it determines the length scale of the
non-Newtonian potential. The potential in Eq.~(\ref{Vy}) can be
positive or negative, depending on the spin of the exchanged boson.
To ease discuss we omit the $\pm$ sign in the following formulas.

To apply the above Yukawa potential to finite nuclei calculation, we
need to do Hartree-Fock calculation as for the Skyrme interaction.
The additional potential energy due to the existence of the Yukawa
potential can thus be written as
\begin{equation}\label{eub}
E_{\rm Y} = \frac{1}{2} \sum_{i,j} \langle ij|V_{\rm Y} (1-P_r
P_\sigma P_\tau) |ij\rangle,
\end{equation}
where $P_r$, $P_\sigma$, and $P_\tau$ are the space, spin, and
isospin exchange operator, respectively, and $|i\rangle$ is the
quantum state of the $i$th particle containing spatial, spin, and
isospin parts. The coefficient $\frac{1}{2}$ is due to the double
counting for exchanging $i$ and $j$.

The first term in Eq.~(\ref{eub}) is the direct (Hartree)
contribution. By expressing the quantum state in the spatial
coordinate representation it is written as
\begin{equation}\label{EUBD}
E_{\rm Y}^D = \frac{1}{2} \int \rho(\vec{r}_1) \rho(\vec{r}_2)
\frac{g^2}{4\pi} \frac{e^{-\mu
|\vec{r}_1-\vec{r}_2|}}{|\vec{r}_1-\vec{r}_2|} d^3r_1 d^3r_2,
\end{equation}
where
\begin{equation}
\rho(\vec{r})=\sum_{i,\sigma,\tau} \phi_{\tau
i}^\star(\vec{r},\sigma)\phi_{\tau i}(\vec{r},\sigma)
\end{equation}
is the nucleon number density with $\phi_{\tau i}(\vec{r},\sigma)$
being the spatial wave function of the $i$th particle with spin
$\sigma$ and isospin $\tau$.

The second term in Eq.~(\ref{eub}) is the exchange (Fock)
contribution. By using $P_\sigma =
(1+\vec{\sigma}_1\cdot\vec{\sigma}_2)/2$ where $\vec{\sigma}_{1(2)}$
is the Pauli operator acting on the state $|i\rangle$ ($|j\rangle$),
it can be written as
\begin{equation}\label{EUBE}
E_{\rm Y}^E = -\frac{1}{4} \sum_{\tau=n,p} \int
[\rho_\tau(\vec{r}_1,\vec{r}_2)\rho_\tau(\vec{r}_2,\vec{r}_1) +
\vec{\rho_\tau}(\vec{r}_1,\vec{r}_2) \cdot
\vec{\rho_\tau}(\vec{r}_2,\vec{r}_1)] \frac{g^2}{4\pi} \frac{e^{-\mu
|\vec{r}_1-\vec{r}_2|}}{|\vec{r}_1-\vec{r}_2|} d^3r_1 d^3r_2,
\end{equation}
where
\begin{equation}
\rho_\tau(\vec{r}_1,\vec{r}_2) = \sum_{i,\sigma}\phi_{\tau
i}^\star(\vec{r}_1,\sigma) \phi_{\tau i}(\vec{r}_2,\sigma)
\end{equation}
and
\begin{equation}
\vec{\rho}_\tau(\vec{r}_1,\vec{r}_2) = \sum_i
\sum_{\sigma,\sigma^\prime}\phi_{\tau
i}^\star(\vec{r}_1,\sigma^\prime) \phi_{\tau i}(\vec{r}_2,\sigma)
\langle \sigma^\prime| \vec{\sigma}|\sigma \rangle
\end{equation}
are the off-diagonal scalar and vector part of the density matrix,
respectively.

The finite-range direct contribution can generally be treated
exactly, while the finite-range exchange contribution has to be
treated using certain approximation. To get similar density
functional form as that from the zero-range Skyrme interaction, we
use in the following calculation the density-matrix expansion
method~\cite{Neg72,Xu10b}. By introducing the coordinate
transformation $\vec{r} = (\vec{r}_1+\vec{r}_2)/2$ and $\vec{s} =
\vec{r}_1-\vec{r}_2$, the off-diagonal scalar and vector part of the
density matrix can be expanded in the lower orders as
\begin{equation}\label{DME1}
\rho_\tau(\vec{r}+\frac{\vec{s}}{2},\vec{r}-\frac{\vec{s}}{2})
\approx \rho_{SL}(k_\tau s) \rho_\tau(\vec{r}) + g(k_\tau s) s^2
\left[ \frac{1}{4} \nabla^2 \rho_\tau(\vec{r}) - \tau_\tau(\vec{r})
+ \frac{3}{5}k_\tau^2 \rho_\tau (\vec{r}) \right]
\end{equation}
and
\begin{equation}\label{DME2}
\vec{\rho}_\tau(\vec{r}+\frac{\vec{s}}{2},\vec{r}-\frac{\vec{s}}{2})
\approx \frac{i}{2}j_0(k_\tau s) \vec{s} \times
\vec{J}_\tau(\vec{r}).
\end{equation}
In the above, $k_\tau = (3\pi^2 \rho_\tau)^{1/3}$ is the Fermi
momentum, and $\tau_\tau$ and $\vec{J}_\tau$ are the kinetic density
and the spin-current density, which can be respectively expressed as
\begin{equation}
\tau_\tau(\vec{r})=\sum_{i,\sigma}|\nabla \phi_{\tau
i}(\vec{r},\sigma)|^2,
\end{equation}
and
\begin{equation}
\vec{J}_\tau(\vec{r})=-i\sum_i \sum_{\sigma,\sigma^\prime}\phi_{\tau
i}^\star(\vec{r},\sigma^\prime) \nabla\phi_{\tau i}(\vec{r},\sigma)
\times \langle \sigma^\prime| \vec{\sigma}|\sigma \rangle.
\end{equation}
The function $\rho_{SL}(k_\tau s)$ and $g(k_\tau s)$ can be
expressed respectively as
\begin{equation}
\rho_{SL}(k_\tau s) = \frac{3j_1(k_\tau s)}{k_\tau s},
\end{equation}
and
\begin{equation}
g(k_\tau s) = \frac{35 j_3 (k_\tau s)}{2(k_\tau s)^3},
\end{equation}
where $j_1$ and $j_3$ are the first- and the third-order spherical
Bessel function.

Applying Eqs.~(\ref{DME1}) and (\ref{DME2}) to Eq.~(\ref{EUBE}), a
Skyrme-like potential energy density functional form from the
exchange contribution of the Yukawa potential can be obtained with
density-dependent coefficients as
\begin{eqnarray}
H_{\rm Y}^E(\vec{r}) = \sum_{\tau=n,p} \{ A[\rho_\tau(\vec{r})] +
B[\rho_\tau(\vec{r})]\tau_\tau(\vec{r}) + C[\rho_\tau(\vec{r})]
[\nabla \rho_\tau(\vec{r})]^2 +
\varphi[\rho_\tau(\vec{r})]J_\tau^2(\vec{r}) \},
\end{eqnarray}
where the expressions of the coefficients are
\begin{eqnarray}
A(\rho_\tau) &=& - \frac{1}{4} \int \rho_\tau^2 \rho_{SL}^2(k_\tau
s) V_{\rm Y}(s) d^3 s
- \frac{3}{5} (3\pi^2)^{2/3} \rho_\tau^{5/3} B(\rho_\tau),\label{a}\\
B(\rho_\tau) &=& \frac{1}{2} \int \rho_\tau \rho_{SL}(k_\tau s)
g(k_\tau
s) s^2 V_{\rm Y}(s) d^3 s,\label{b}\\
C(\rho_\tau) &=& \frac{1}{4}\frac{dB(\rho_\tau)}{d\rho_\tau},\label{c}\\
\varphi(\rho_\tau) &=& -\frac{\pi}{6} \int j_0^2(k_\tau s) s^4
V_{\rm Y}(s) ds,
\end{eqnarray}
with $V_{\rm Y}(s) =  g^2e^{-\mu s}/(4\pi s)$.

In the standard Skyrme-Hartree-Fock calculation, the finite nuclei
properties are obtained by solving the following Schr\"{o}dinger
equation
\begin{eqnarray}
\left[-\nabla \cdot \frac{\hbar^2}{2m_\tau^\star(\vec{r})} \nabla +
U_\tau(\vec{r}) + \vec{W}_\tau (\vec{r}) \cdot (-i\nabla \times
\vec{\sigma})\right] \phi_{\tau i} = e_{\tau i} \phi_{\tau i},
\end{eqnarray}
where $e_{\tau i}$ is the eigenvalue of the single-particle energy
and $\phi_{\tau i}$ is the nucleon wave function. Due to the
existence of the Yukawa-type non-Newtonian gravitational potential,
the single-particle potential $U_\tau$, the nucleon effective mass
$m_\tau^\star$, and the spin-orbit potential $\vec{W}_\tau$ are not
only determined by the Skyrme interaction but modified by the Yukawa
potential as well, and they can be calculated from the variational
principle. The modifications of the single-particle potential from
the Yukawa potential are from both the direct contribution and the
exchange contribution
\begin{eqnarray}
U_\tau^{\rm Y} &=& U_\tau^{\rm YD} + U_\tau^{\rm YE}, \\
U_\tau^{\rm YD} &=& \int \rho(\vec{r}^\prime) \frac{g^2}{4\pi}
\frac{e^{-\mu
|\vec{r}-\vec{r}^\prime|}}{|\vec{r}-\vec{r}^\prime|} d^3 r^\prime, \label{d} \\
U_\tau^{\rm YE} &=& \frac{dA(\rho_\tau)}{d\rho_\tau} +
\frac{dB(\rho_\tau)}{d\rho_\tau}\tau_\tau -
\frac{dC(\rho_\tau)}{d\rho_\tau} (\nabla \rho_\tau)^2 -
2C(\rho_\tau) \nabla^2 \rho_\tau +
\frac{d\varphi(\rho_\tau)}{d\rho_\tau} J_\tau^2. \label{ex1}
\end{eqnarray}
We note that the direct contribution is much larger than the
exchange contribution. The nucleon effective mass and the spin-orbit
potential are modified only by the exchange contribution of the
Yukawa potential, respectively, according to
\begin{equation}
\frac{\hbar^2}{2m_\tau^\star} \rightarrow
\frac{\hbar^2}{2m_\tau^\star} + B(\rho_\tau) \label{ex3}
\end{equation}
and
\begin{equation}
\vec{W}_\tau^{\rm Y}=2\varphi(\rho_\tau)\vec{J}_\tau. \label{ex2}
\end{equation}

\section{Results and discussions}

Using the formulism discussed in Sec.~\ref{theory}, we are now able
to study the properties of finite nuclei in the presence of the
Yukawa-type non-Newtonian gravitational potential. In the following,
we will discuss the effects from the Yukawa potential by comparing
its strength with the Coulomb interaction in finite nuclei and then
set an upper limit of its strength.

\subsection{Comparing with the Coulomb interaction}

\begin{figure}[h]
\begin{center}
\includegraphics[scale=1.0]{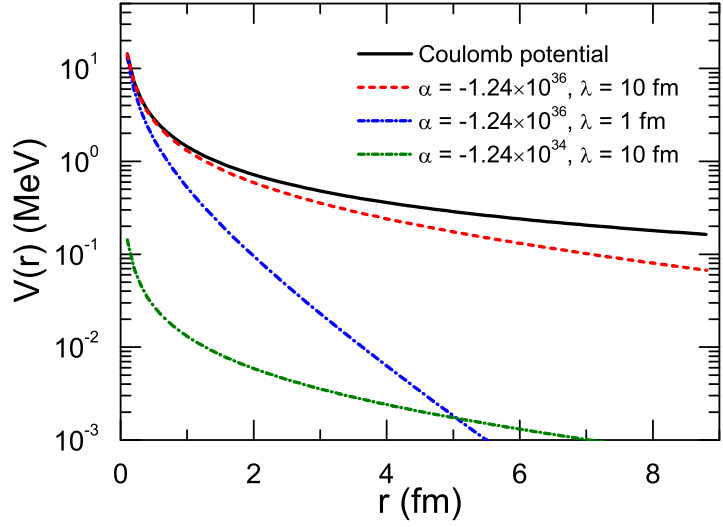}
\end{center}
\caption{\label{vr} Comparing the Coulomb potential with the
non-Newtonian Yukawa potential between two protons of different
strength and length scale parameters at different distances of
separation. Taken from Ref.~\cite{Xu12}.}
\end{figure}

It is instructive to set the strength of the Coulomb interaction as
a base line and discuss how the finite nuclei properties will be
affected if the non-Newtonian potential in finite nuclei is as
strong as the Coulomb potential. In Fig.~\ref{vr} we compare the
Coulomb potential with the non-Newtonian Yukawa potential between
two protons with $\alpha=-1.24\times10^{36}$ or
$-1.24\times10^{34}$, and $\lambda=1$ fm or 10 fm, respectively. We
note that $\alpha=-1.24\times10^{36}$ leads to the coupling constant
$g^2/4 \pi = 1/137$, which is exactly the same strength as the
Coulomb potential. Thus, at short distances the Yukawa potential is
similar to the Coulomb potential while it decreases faster for a
smaller value of $\lambda$ due to the exponential decay term. The
Yukawa potential with only $1\%$ strength of that is also shown for
reference.

\begin{figure}[h]
\begin{center}
\includegraphics[scale=1.0]{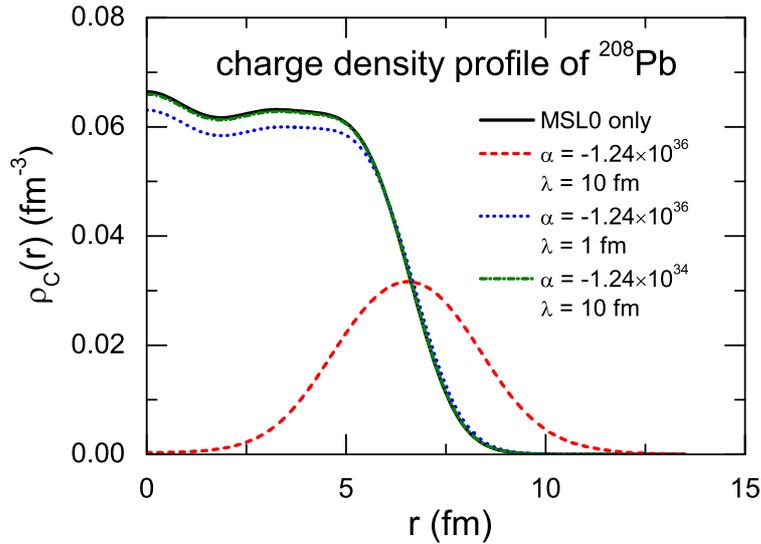}
\end{center}
\caption{\label{den} Charge density profiles of lead nucleus from
Skyrme-Hartree-Fock calculation with the MSL0 Skymre force only and
with an additional non-Newtonian Yukawa potential of different
strength and length scale parameters. Taken from Ref.~\cite{Xu12}.}
\end{figure}

Next, let's see what would happen in finite nuclei if the Yukawa
potential is as strong as the Coulomb potential. It is seen in
Fig.~\ref{den} that the charge density profile of lead nucleus will
be more diffusive in the presence of the repulsive Yukawa potential,
compared to the result from MSL0 force only. Thus, the repulsive
Yukawa potential as strong as the Coulomb potential would generally
increase the charge radii of finite nuclei, and the effect is larger
with a larger $\lambda$. This is quite understandable, and it is
seen that the effect becomes negligible if the strength of Yukawa
potential is $1\%$ of the Coulomb potential.

\subsection{Nuclear Constraint on the Non-Newtonian potential}

From the above discussion, we intuitively know that the Yukawa-type
non-Newtonian gravitational potential will change the finite nuclei
properties unless it is of short range and/or much weaker than the
Coulomb potential. To avoid modifying the well-determined
nucleon-nucleon interaction available, we require that the Yukawa
potential will not change much the finite nuclei properties which
have been well described by our framework in nuclear physics, at
least within the uncertainty from nuclear predictions. Especially,
the charge radii and binding energies, which can be very accurately
measured in experiments, are mostly related to the well-constrained
isoscalar part of the nucleon-nucleon interaction but are not
sensitive to the less certain isovector part and/or three-body
interaction, and they can now be reproduced by the MSL0 Skyrme force
within about $2\%$. This $2\%$ uncertainty is thus the largest room
for the possible existence of the non-Newtonian gravity at
femtometer scale.

\begin{figure}[h]
\begin{center}
\includegraphics[scale=1.0]{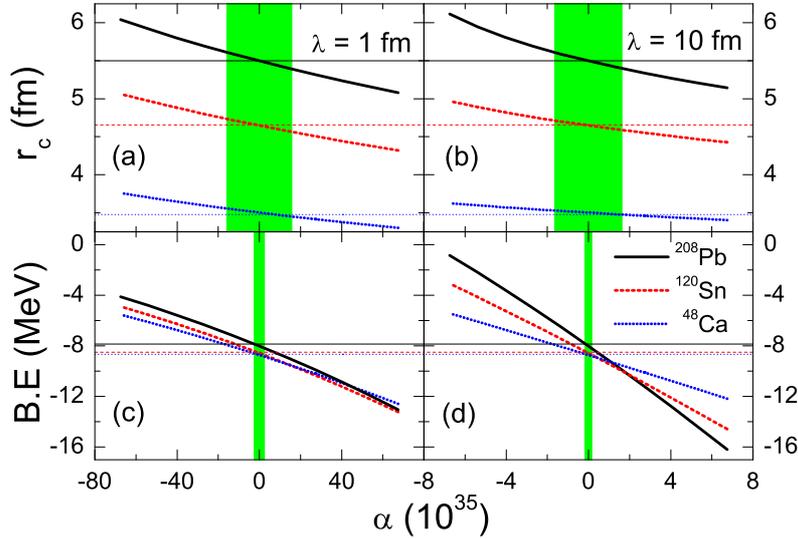}
\end{center}
\caption{\label{properties} The charge radii ($r_c$) and binding
energies per nucleon (B.E.) of $^{208}$Pb, $^{120}$Sn, and $^{40}$Ca
nuclei as functions of the strength parameter $\alpha$ of the
non-Newtonian Yukawa potential at length scales $\lambda=1$ and 10
fm. The horizontal lines are the mean values of the experimental
data~\cite{Aud03,Ang04}. $\alpha=0$ is the case without
non-Newtonian potential, and the green band is the largest room for
$\alpha$ that the change of the charge radius or binding energy of
$^{208}$Pb due to the existence of the non-Newtonian potential is
within $2\%$. Taken from Ref.~\cite{Xu12}.}
\end{figure}

How the charge radii and binding energies from medium to heavy
nuclei change with the strength parameter $\alpha$ of the Yukawa
potential at length scales $\lambda=1$ and 10 fm is displayed in
Fig.~\ref{properties}. In the presence of the repulsive (attractive)
Yukawa potential, corresponding to a negative (positive) value of
$\alpha$, the charge radii increase (decrease) and the nuclei become
less (more) bound, which is quite understandable. The change is
almost linear to $\alpha$ and more sensitive to the strength
parameter for $\lambda=10$ fm than for $\lambda=1$ fm. In addition,
the change is larger for heavy nuclei $^{208}$Pb due to the
finite-range nature of the Yukawa potential. Also plotted in the
figure is the $2\%$ uncertainty of both charge radius and binding
energy of $^{208}$Pb from the MSL0 force only, which are shown with
green bands. The green bands are thus the constraints for the
strength parameter of the Yukawa potential.

\begin{figure}[h]
\begin{center}
\includegraphics[scale=1.0]{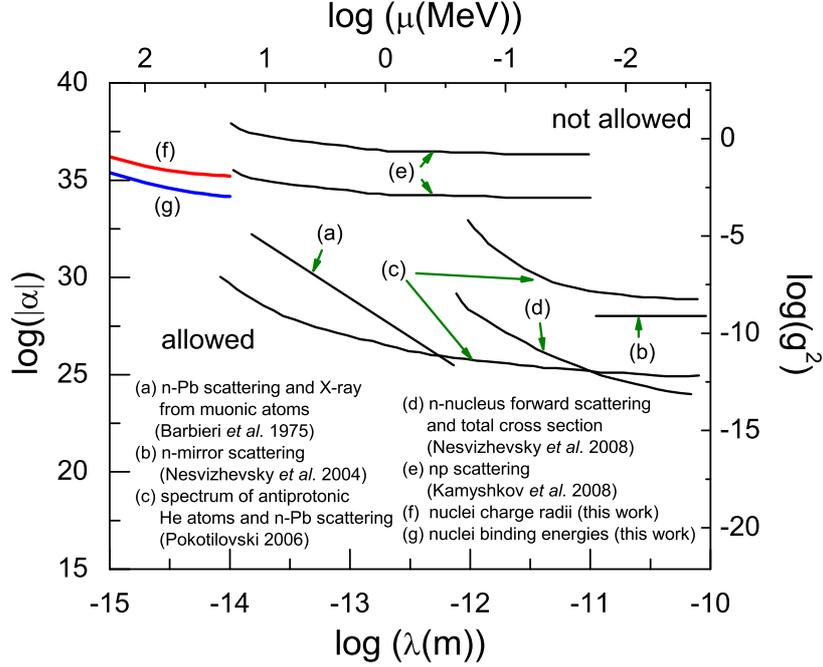}
\end{center}
\caption{\label{constraint} Constraints of the non-Newtonian Yukawa
potential at femtometer scale from nuclei binding energies and
charge radii in the present work together with those from neutron
scattering experiments at larger scales. Taken from
Ref.~\cite{Xu12}.}
\end{figure}

The constraints on the strength of the non-Newtonian potential at
femtometer scale from nuclei binding energies and charge radii
obtained above are compared with those from other works mostly using
neutron scattering experiments~\cite{Bar75,Nes04,Pok06,Nes08,Kam08}
at larger scales in Fig.~\ref{constraint}. It is found the upper
limit is larger at shorter distances, reflecting the increasing
difficulty of the experiments and the stronger potential at work at
smaller scale. In addition, it is found that our constraint covers
the previously unexplored region and is a smooth expansion of those
at larger scales. Typically, the constraint from nuclei binding
energy can be written as
\begin{equation}
\log(|\alpha|)<1.75/[\lambda(\rm fm)]^{0.54} + 33.6,
\end{equation}
and that from nuclei charge radius can be written as
\begin{equation}
\log(|\alpha|)<1.18/[\lambda(\rm fm)]^{0.79} + 35.0,
\end{equation}
for $\lambda=1 \sim 10$ fm. Since the constraint from the binding
energy is stronger, it can be used as the nuclear constraint on the
non-Newtonian gravity at femtometer scale, and it also leads to the
upper limit of the boson-nucleon coupling constant in the form of
\begin{equation}
\log(g^2)<0.10 [\mu (\rm MeV)]^{0.54}-3.53,
\end{equation}
if the boson mass $\mu$ is between 20 MeV and 200 MeV.

\section{Summary and Outlook}

In this talk we report our recent results of nuclear constraints on
non-Newtonian gravity at femtometer scale. To do this, we
consistently incorporate the Yukawa-type non-Newtonian gravitational
potential in the Skyrme-Hartree-Fock calculation. The nucleon
interaction is represented by the well-established MSL0 Skyrme
force. By assuming that the Yukawa potential will not change the
binding energies and charge radii of finite nuclei by $2\%$, the
strength of the Yukawa potential is constrained within
$\log(|\alpha|)<1.75/[\lambda(\rm fm)]^{0.54} + 33.6$ for $\lambda=1
\sim 10$ fm, so that the calculated properties of finite nuclei will
not be in conflict with the very accurate experimental data
available. This constraint may serve as a useful reference in
constraining properties of weakly-coupled gauge bosons and further
explorations of possible extra dimensions at femtometer scale.

\ack

This work was supported in part by the US National Science grants
PHY-0757839 and PHY-1068022, the National Aeronautics and Space
Administration under grant NNX11AC41G issued through the Science
Mission Directorate, the National Natural Science Foundation of
China under Grant Nos. 10975097 and 11135011, Shanghai Rising-Star
Program under grant No. 11QH1401100, ¡°Shu Guang¡± project supported
by Shanghai Municipal Education Commission and Shanghai Education
Development Foundation, the Program for Professor of Special
Appointment (Eastern Scholar) at Shanghai Institutions of Higher
Learning, the Science and Technology Commission of Shanghai
Municipality (11DZ2260700), and the National Basic Research Program
of China (973 Program) under Contract No. 2007CB815004.

\end{document}